# Microscopic origin of light emission in $Al_yGa_{1-y}N$/GaN superlattice: Band profile and active site

Duanjun Cai[1], Junyong Kang [2], and Guang-Yu Guo [1, 3*]

[1] Department of Physics, National Taiwan University, Taipei 106, Taiwan

[2] Department of Physics and Semiconductor Photonics Research Center, Xiamen University, Xiamen 361005, China

[3] Graduate Institute of Applied Physics, National Chengchi University, Taipei 116, Taiwan

We present first-principles calculations of AlGaN/GaN superlattice, clarifying the microscopic origin of the light emission and revealing the effect of local polarization within the quantum well. Profile of energy band and distributions of electrons and holes demonstrate the existence of a main active site in the well responsible for the main band-edge light emission. This site appears at the position where the local polarization becomes zero. With charge injection, the calculated optical spectra show that the broadening of the band gap at the active site leads to the blue shift of emission wavelength.

Over the past decade, GaN and its related alloys have attracted enormous interest because of their important application in short-wavelength and high-power optoelectronic devices. The first generation of nitride-based light emitting devices was based on the ternary InGaN compound as the high efficient active layers [1, 2]. As an inevitable trend for exploiting shorter wavelength and broader band, the coming generation of light devices naturally turns to the AlGaN system that possesses wider energy band gap up to about 6.2 eV [3]. In practice, the strong polarization and large dislocation concentration in wurtzite nitride are two important problems associated with the improvement of the luminescence efficiency. Fortunately, the high-brightness luminescence has been achieved in the InGaN based light emitting diodes (LEDs), which benefits from the remarkable compositional inhomogeneities of InN quantum dots [4] and the dislocation reducing techniques [5, 6]. In such case, the problems mentioned above were skillfully avoided to some extent. However, the AlGaN system has a much larger polarization constant [7] and does not show those special properties of inhomogeneities. Therefore, to gain high emission efficiency in the AlGaN system is a challenge associated with controlling the impacts of polarization.

In modern technology, quantum well and superlattice is frequently used as the active core layers in the optoelectronic components, in virtue of the quantum confinement effect. However, the piezoelectric polarization in the quantum well meanwhile becomes much stronger and complicated [8, 9]. Recently, breakthroughs in the AlGaN superlattice based LEDs have been achieved in the ultraviolet to ultra-deep ultraviolet waveband though the emission efficiency remains low [10–12]. Important issues such as band offset, misfit strain, electron confinement, and luminescence spectra, on the AlN/GaN superlattice were studied by many groups using either experimental or theoretical methods [13 –16]. Although many achievements have been made, because the structure of superlattice is in the atomic or nanometer scale, the microscopic origin of the light generation in the AlGaN/GaN remains ambiguous and is desired in order to accurately control and further develop nitride-based light emitting devices.

In this letter, we reveal the existence of the active site within the AlGaN/GaN quantum well responsible for light emission and demonstrate the influence of site-dependent polarization by *ab initio* calculations. The profile of band configuration and the distribution of carriers (electrons and holes) of AlGaN/GaN superlattice is obtained. We find that the highest electron-hole (*e-h*) overlap occurs at the site where the local polarization becomes zero. From optical spectra, this site is identified to be the main active site associated with the main band-edge light emission. With injection of various charge densities to the superlattice, blue shift of light wavelength is observed deriving from the broadening of the local band gap of the active site, suggesting that the current density can be used to accurately tune the main wavelength of light.

Our approaches were based on first-principles calculations in the framework of density functional theory [17], using the Vienna *ab initio* simulation package (VASP) [18] and also home-made codes for calculating optical properties [19]. The interaction between ions and electrons was described by the projector augmented wave (PAW) method [20] and the plane-wave cutoff energy was set to 500 eV. The supercell of $Al_{0.5}Ga_{0.5}N$/GaN superlattice along the [0001] direction was constructed with barrier and well width each 2.2 nm, comparable with the size of experimental quantum well structure in optoelectronic devices. The atomic structure was fully relaxed. The AlN fraction in $Al_{0.5}Ga_{0.5}N$ barrier was evenly distributed for the homogeneous chemical composition. It is well known that DFT does not describe the optical properties of semiconductors very well due to its neglect of, .e.g, quasiparticle self-energy corrections [21]. In this letter the scissors operator technique was used to remove the band gap problem [22].

Structural optimizations show that the $Al_{0.5}Ga_{0.5}N$ barrier layers largely relax towards its equilibrium constant and the residual in-plane strain is nearly zero, whereas the GaN well layers remains under a compressive strain of -1.5%. It reflects that AlN layers prefer to stay in a stress-free state while GaN possesses large tolerance to accommodate misfit strain. This is because AlN has a stronger polarity than GaN does, thus, imposing an external stress on AlN would cost considerable energy [23]. This is consistent with the fact that the shorter-wavelength LEDs using AlGaN/GaN active layers have been achieved based on the AlN or AlGaN basal layers rather than on the GaN one [11, 12]. On the other hand, Fig.1 shows the the variation of the c-axis length of cells within the GaN quantum well. The equilibrium c-axis length of unstrained GaN unit cell is 5.184 angstroms. Therefore, one can see that the local strain is inhomogeneous along the [0001] direction. In the GaN quantum well, the c-axial strain close to interface 1 increases whereas the strain in the side near interface 2 remains constant. Similarly, non-uniform lattice distortion of local structures in the nanoparticles has been observed experimentally [24]. This inhomogeneous local strain would influence the local band gap, as discussed below.

In general, the shape of potential well is widely regarded as the fundamental information for quantum well and superlattice. Although several experimental and theoretical methods have been developed to describe this potential well [7, 16, 25], the complete configuration of the actual potentials for electrons and holes in the quantum well of semiconductor system is still difficult to obtain. Here we first calculated the bi-layer decomposed density of states (DOS), and then, by joining the edges of the bi-layer [Ga(Al) and N] local DOS of conduction (CB) and valence (VB) bands along the [0001] direction, we obtained the profile of band configuration of the quantum well and barrier in the $Al_{0.5}Ga_{0.5}N$/GaN superlattice (Fig. 2). From this profile, three important features of the CB and VB potential wells can be clearly identified: i) Both configurations of the barrier and well of the CB and VB experience a bending from the flat band. This is due to the effect of the spontaneous and piezoelectric polarizations and their differences between AlGaN and GaN system; ii) The bending turns out to be more complicated than a widely accepted simple linear tilting. The CB well forms a potential valley with the minimum in the left side near interface 1, while the VB well have the minimum on the right side close to interface 2. In fact, this may result from the inhomogeneity of the stress field within the GaN well as mentioned above; iii) The bending for CB and VB is rather different and asymmetric. The average band offset of the CB of 0.72 eV is larger than that of the VB of 0.44 eV. Furthermore, the CB is significantly curved and the well is deep whereas the VB well appears flat, especially in the right side near interface 2.

From the above discussion, it is clear that the polarization effect on the band bending in the nitride heterostructure is more complicated than the simple description assuming only the uniform electric field. The difference in the band curvature and the potential depth between CB and VB reflects their different capabilities in trapping carriers and also, the different displacement rates of electrons and holes by the polarization field. In other words, the spatial *e-h* overlap within the quantum well may be much higher than that assuming only the symmetric CB and VB bending. To clarify this important point, the real-space distributions of electrons at the conduction band minimum (CBM) and holes at the valence band maximum (VBM) were displayed in Fig. 3. One can see that electrons and holes are entirely confined in the quantum well and displaced away from the well center by the effect of polarization *P*. The electrons shift towards interface 1 and have the maximal accumulation at about 8.5 Å (layer $R_0$) from interface 1. In contrast, the holes reside in the right part of the GaN well near interface 2 and distribute widely. In this different degree of polarization, electron and hole envelopes still overlap effectively with each other in the GaN well and the maximum overlap rate occurs at $R_0$. In principle, the higher overlap rate would result in the higher probability of the *e-h* recombination. In details, Fig. 3(c-d) show that the CBM electrons come mainly from the *s*-like states whereas the VBM holes derive from the $p_x$-like states. Importantly, these electrons and holes are tightly bound to the N atoms. This indicates that the N atoms in the GaN quantum well act as the main recombination center for the light emission, which contributes to the on-site transition of *e-h* recombination. It is known that the probability of on-site transition, compared to that of off-site transitions, dominates the total transitions. Therefore, the local interaction of electrons and holes at this center would directly determine the light emission efficiency.

Comparing the *s*-states electrons at different sites in Fig. 3(c), we find, surprisingly, that the electron cloud deforms from the spherical shape and its charge center displaces from the site of the nitrogen nuclei. The direction and distance of the center displacement appears strongly site-dependent. This fact demonstrates the influence by the local polarization rather than the previously discussed uniform polarization field, which should have induced homogeneous displacement. It is known that the emission properties in GaN/AlGaN heterostructures are often affected by such localization effects as alloy disorder, interface roughness, or structural defects [4, 26]. However, the effect of local polarization observed here suggests the microscopic nature of light emission in nitride heterostructure systems. Fig. 4 shows the charge center displacement as a function of the position of nitrogen nuclei within the GaN well, reflecting the local polarization. The displacement shifts to the [000-1] (negative) and [0001] (positive) directions on the sides towards interface 1 and 2, respectively. In particular, at site $R_0$ the local polarization is eliminated and the displacement become zero. Since the local polarization on the $p_x$-state holes is negligibly small [Fig. 3(d)], the displacement of $s$-electrons decides the spatial overlap rate of local $e$-$h$ bound to the N atoms. Therefore, the maximal effective $e$-$h$ overlap would occur near the site $R_0$ of zero-polarization and this site is the main active site for the light emission in quantum well.

In order to understand the correlation between the spectral peak and the generating site, the optical spectrum was calculated by using Fermi golden rule [19]. The local field effect in our system is found to be small by using the Maxwell-Garnett formula [27]. The absorptive (imaginary) part of the dielectric constant ($\varepsilon''$) of bulk GaN and $Al_{0.5}Ga_{0.5}N$/GaN superlattice are plotted in Fig. 5a. The $\varepsilon''$ of bulk GaN calculated within the DFT+scissor-operator approach agrees well with the experimental spectrum [28], especially near the band edge. The light emission rate $R_{vc}$ can be reflected from $\varepsilon''$ as described by the van

Roosbroek-Shockley relation [29]: $R_{vc} = \rho \varepsilon''$, where $\rho$ is the photon density that is given by $8\pi v^2 n^3 / c^3$. One can see that near the band edge, the emission intensities ($z_1 \sim z_4$) of $Al_{0.5}Ga_{0.5}N$/GaN superlattice are significantly enhanced due to the quantum confinement of the GaN quantum well. Detailed analysis of the transition matrix elements and the involved transition levels at different sites show that the intense band-edge peak $z_1$ comes mainly from the $\Gamma_1$-$\Gamma_{6'}$ transition at $R_0$ site (Fig. 5b). The DOS at the $\Gamma_1$ and $\Gamma_{6'}$ levels are found to be the highest compared to those at other levels in CB and VB. The peaks at higher energy ($z_2 \sim z_4$) are also assigned to the transitions generated from $R_0$. Therefore, it can be confirmed that the site $R_0$ in the GaN quantum well, with a maximal *e-h* overlap, is the most active light emission center. In light of this fact, since the active site is highly localized, a light source for lasers with sharp monochromatic light emission can be better achieved by inserting additional restriction layers enclosing the active site.

In addition, the peak energy ($z_1$) of AlGaN/GaN superlattice system is found to have a red shift from that of the bulk GaN, due to the quantum confinement Stark effect [30]. To further understand this effect as well as the charging evolution of polarization and light emission under LED working conditions, different electron concentrations $N_e$ ($10^{17} \sim 10^{20}$ cm$^{-3}$) were injected into the superlattice. As shown in Fig. 5a, the charge injection considerably enhances the spectral intensity and bring about a blue shift of the $z_1$ peak. The phenomenon of blue shift has also been observed in the luminescence spectra of InN-based LEDs by increasing working current [31]. We clarify the origin of these effects as follows: a) The spectral shift results from the changes of the effective band gap $E_g^{eff}$, the band gap at the main active site ($R_0$) for the main band-edge emission (Fig. 2). Table I shows that $E_g^{eff}$ in Al$_{0.5}$Ga$_{0.5}$N/GaN I is narrower than that of bulk GaN, due to the band bending and the enhancement of polarization. However, the $E_g^{eff}$ is broadened by the charge injection (Al$_{0.5}$Ga$_{0.5}$N/GaN II and III), which results in the blue shift. b) The enhanced emission efficiency reflects the increased overlap of the local $e$ and $h$ envelopes at the active site (Fig. 3). This envelope deformation is caused by the partial screening of polarization with charge injection (Table I). When $N_e$ exceeds $10^{19}$ cm$^{-3}$, both the CB and VB potential quantum wells become centrally symmetric and the $e$-$h$ overlap rate at the active site increases significantly. This explains at the microscopic level why the operation of nitride-based laser diodes would need a rather high threshold carrier density of $10^{19}$ cm$^{-3}$ [9, 32].

The authors acknowledge supports from National Science Council and NCTS of Taiwan as well as TSMC under TSMC JDP #NTU-0806 and NCTS.

# References

\* Electronic address: gyguo@phys.ntu.edu.tw

FIGURE CAPTIONS:

**FIG. 1.** (color online) The c-axis length of unit cell within the GaN quantum well. This shows the local strain in different regions.

**FIG. 2.** (color online) Profile of band edge configuration of $Al_{0.5}Ga_{0.5}N$/GaN superlattice.

**FIG. 3.** (color) Distribution of carriers in the $Al_{0.5}Ga_{0.5}N$/GaN superlattice. Longitudinal contour of electrons at CBM, (a), and of holes at VBM, (b). (c) and (d), 3-dimensional images of the charge clouds in (a) and (b), respectively. The displacement and deformation of $s$-like electron states in (c) reveals the influences of local polarization.

**FIG. 4.** Displacement of local electron center as a function of position in GaN quantum well. The $R_0$ site with zero local polarization produces the maximum overlap of electrons and holes.

**FIG. 5.** (color online) (a) Absorptive spectra of bulk GaN, $Al_{0.5}Ga_{0.5}N$/GaN superlattice, and the superlattice with charge injection. Shifts of $z_1$ peak at the band edge can be observed. (b) Decomposed electronic band structures at $R_0$ site. Purple dots in different size demonstrate the DOS magnitude therein. The origin of the transitions for the $z_1 \sim z_4$ peaks in (a) are one-to-one identified.

TABLE I. Effective band gap $E_g^{eff}$ and Polarization $P$ of bulk GaN and $Al_{0.5}Ga_{0.5}N$/GaN superlattice under various electron concentrations $N_e$. (R) and (B) represents the red and blue shift, respectively, of the corresponding band-edge peaks.

| | $N_e$ (cm$^{-3}$) | $E_g^{eff}$ (eV) | $P$ (C/m$^2$) |
|---|---|---|---|
| bulk GaN | 0 | 3.42 | -0.027 |
| $Al_{0.5}Ga_{0.5}N$/GaN I | 0 | 3.32 (R) | -0.089 |
| $Al_{0.5}Ga_{0.5}N$/GaN II | $8.7 \times 10^{17}$ | 3.41 (B) | -0.014 |
| $Al_{0.5}Ga_{0.5}N$/GaN III | $8.7 \times 10^{18}$ | 3.45 (B) | -0.012 |

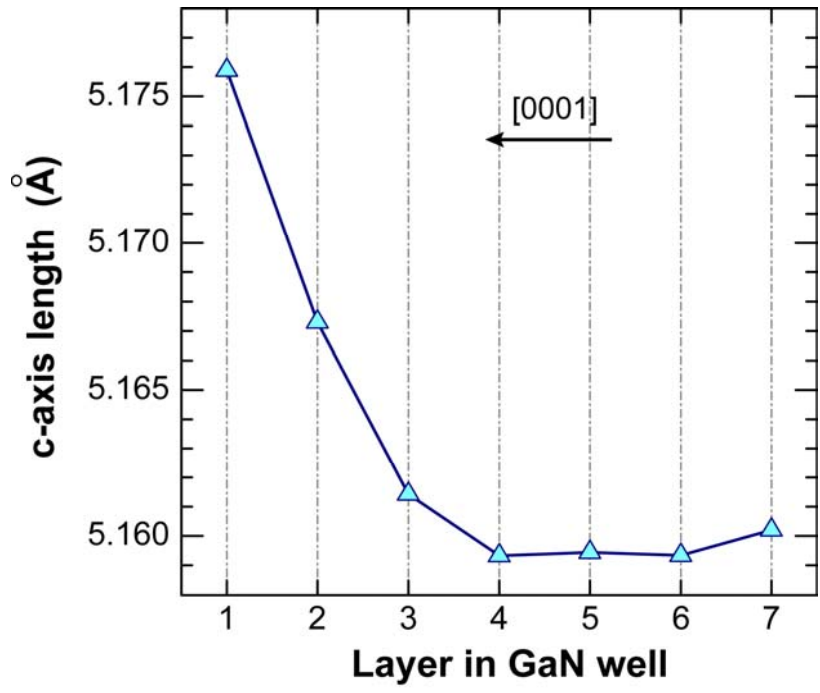

FIG. 1. D. Cai et al.

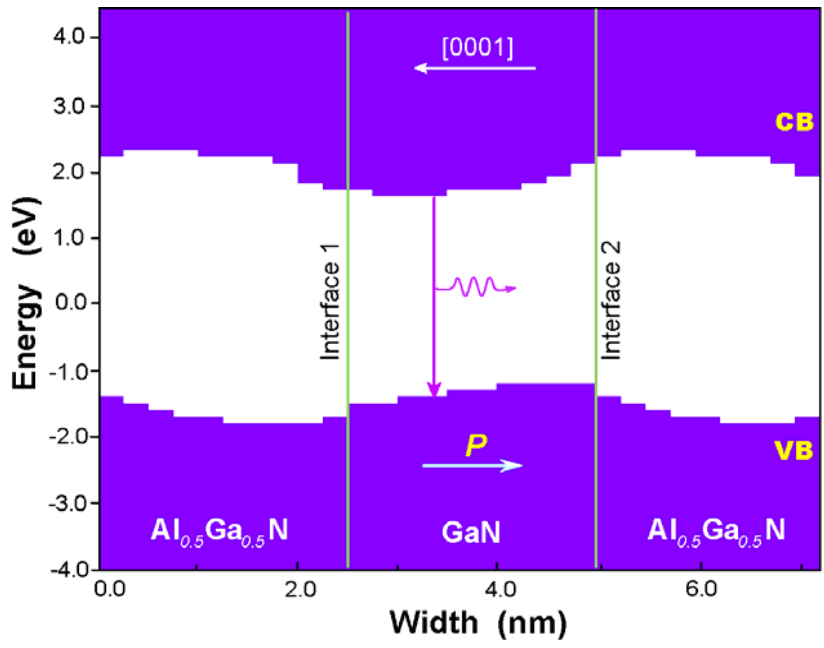

FIG. 2. D. Cai et al.

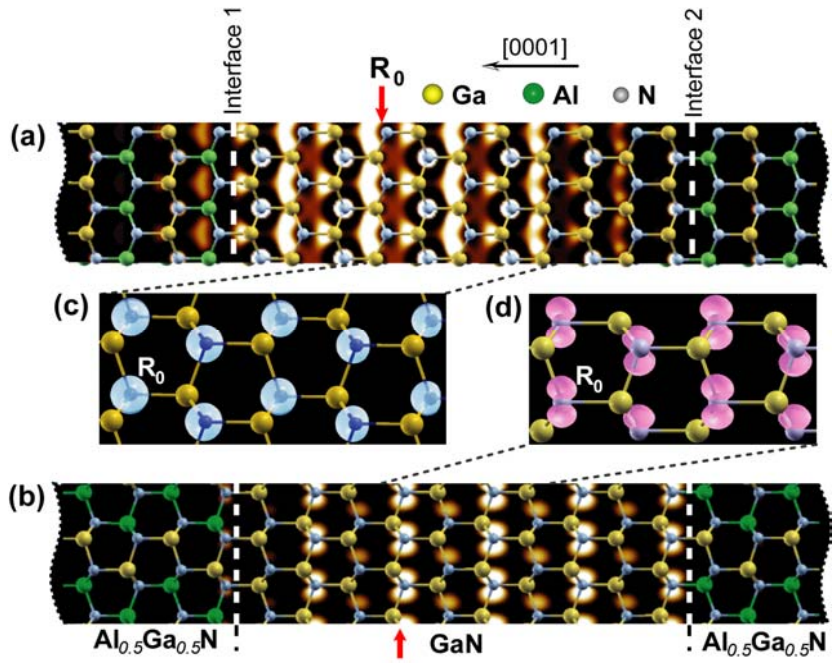

FIG. 3. D. Cai et al.

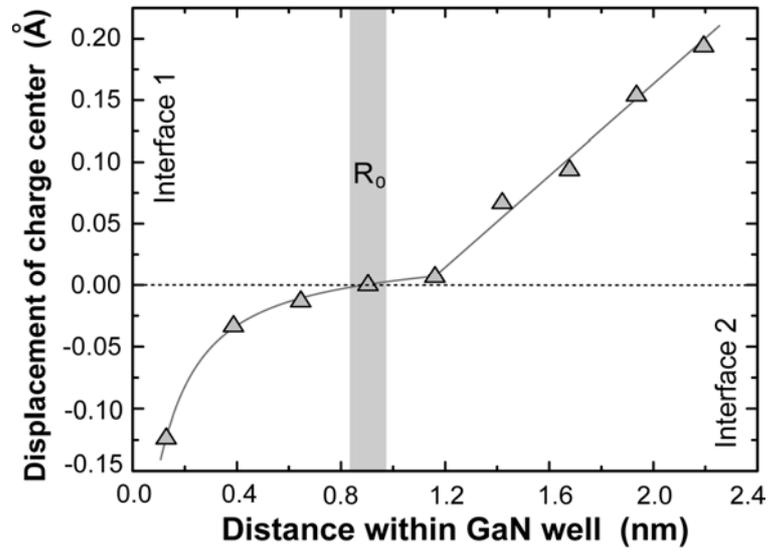

FIG. 4. D. Cai et al.

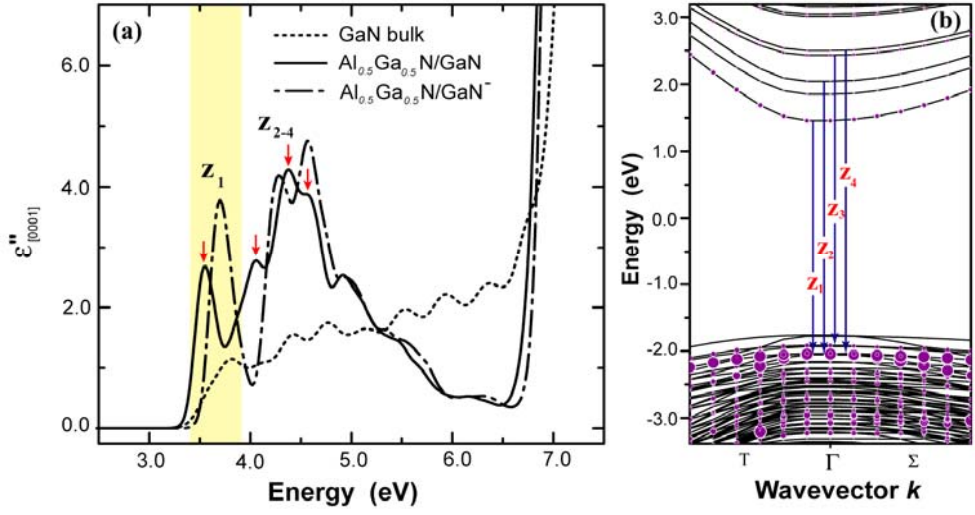

FIG. 5. D. Cai et al.